\documentclass[aps,prl,twocolumn,superscriptaddress,longbibliography]{revtex4-2}
\usepackage[colorlinks=true,bookmarks=true,citecolor=magenta,urlcolor=magenta,linkcolor=magenta,breaklinks]{hyperref} 
\usepackage{breakurl}
\usepackage{xcolor, soul}
\usepackage{sidecap}
\usepackage{amssymb}
\usepackage{hhline}
\usepackage{multirow}
\sidecaptionvpos{figure}{t}
\usepackage{amsmath}
\usepackage{mathrsfs}
\usepackage{graphicx}
\usepackage{esint}
\usepackage{epstopdf}
\usepackage{rotating}
\usepackage{framed}
\graphicspath{{pict/}{./}}
\usepackage{bm}%
\usepackage{microtype,bm,bbm,graphicx,booktabs,times}

\newcounter{Fig}

\newcommand\mymapstol{\mathrel{\ooalign{$\leftarrow$\cr%
  \kern1.75ex\raise0.275ex\hbox{\scalebox{1}[0.4]{$\mid$}}\cr}}}

\newcommand\mymapstor{\mathrel{\ooalign{$\rightarrow$\cr%
  \kern-.15ex\raise.275ex\hbox{\scalebox{1}[0.4]{$\mid$}}\cr}}}
  
\usepackage{graphicx,tipa}

\usepackage{tikz}

\makeatletter
\newcommand*{\rom}[1]{\expandafter\@slowromancap\romannumeral #1@}
\makeatother
\begin{document}
\title{Topological Optical Achirality}
\author{Chunchao Wen}
\author{Zhichun Qi}
\author{Jianfa Zhang}
\email{jfzhang85@nudt.edu.cn}
\author{Chaofan Zhang}
\email{c.zhang@nudt.edu.cn}
\author{Shiqiao Qin}
\author{Zhihong Zhu}
\author{Wei Liu}
\email{wei.liu.pku@gmail.com}
\affiliation{College for Advanced Interdisciplinary Studies, National University of Defense Technology, Changsha 410073, P. R. China.}
\affiliation{Nanhu Laser Laboratory and Hunan Provincial Key Laboratory of Novel Nano-Optoelectronic Information Materials and
Devices, National University of Defense Technology, Changsha 410073, P. R. China.}

\begin{abstract}
For arbitrary reciprocal single-mode structures, regardless of their geometric shapes or constituent materials, there must exist incident directions of plane waves for which they are optically achiral. 
\end{abstract}

\maketitle


Optical chirality for light-matter interactions, characterized generally by distinct optical responses for right and left handedly circularly-polarized [RCP ($\circlearrowright$) and LCP ($\circlearrowleft$)] incident light, is a fundamental concept pervading most branches of photonics~\cite{BARRON_2009__Molecular,MUN_LightSci.Appl._Electromagnetic}. Historical studies on optical chiralities in the $19^{\mathrm{th}}$ century were largely focused on systems consisting of randomly distributed geometrically chiral molecules in large quantities. As a result, the system investigated is effectively isotropic, with its optical responses being independent of incident directions~\cite{BARRON_2009__Molecular}. In other words, the chiroptical responses observed were essentially orientation-averaged (or incident-direction averaged) for the consisting molecule, establishing  the connection between optical and geometric chiralities: geometric achirality (exhibiting structural mirror or inversion symmetry) of the molecule would inevitably lead to optical achirality (identical optical responses for incident RCP and LCP waves) of the whole system, as is required by the law of parity conservation~\cite{BARRON_2009__Molecular}. 

For general photonic structures, optical responses are dependent on incident directions, and so are their optical chiralities~\cite{BARRON_2009__Molecular,MUN_LightSci.Appl._Electromagnetic}. Basically there are no definitive connections between geometric and optical chiralities~\cite{CHEN_Phys.Rev.Lett._Extremize}: for some incident directions, geometrically chiral structures can be optically achiral and geometrically achiral structures can be optically chiral; structure of a fixed geometric chirality can manifest opposite optical chiralities for different incident directions. It is well known that a structure can be optically achiral for all incident directions, such as a metal bar supporting solely an electric dipole~\cite{CHEN_Phys.Rev.Lett._Extremize}. Our central question is: \textit{{Is it possible for a structure to be optically chiral for all incident directions?}} If such a structure exists, it should not exhibit any mirror symmetry, since parity conservation ensures that for all incident directions parallel to the mirror plane it is optically achiral.

Here we provide a partial answer to the above question, showing that for reciprocal single-mode [quasi-normal mode (QNM) for open non-Hermitian systems~\cite{LALANNE__LaserPhotonicsRev._Light}]  structures the answer is impossible. Independent of the optical parameters of the consisting materials and the geometric shape of the structure, as long as it is reciprocal and supports dominantly one single mode, then there must be incident directions of plane waves along which it is optical achiral. Such optical achirality is protected by fundamental principles of reciprocity and global topology, which can find applications in not only chiral optics and topological photonics, but also many other branches of wave physics where reciprocity and topology are generic and ubiquitous.

\begin{figure}[tp]
\centerline{\includegraphics[width=0.4\textwidth]{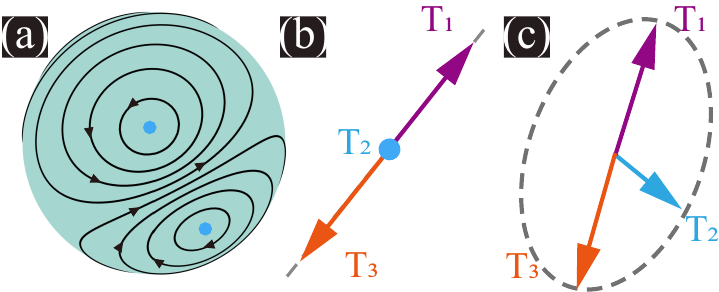}} \caption {\small (a) Vector fields on a sphere with two topologically protected singularities (vectorial zeros), as required by the Poincaré-Hopf theorem. Linear and elliptic polarizations are shown respectively in (b) and (c), for which instantaneous field vectors are shown at three instants ($\mathrm{t=T_{1,2,3}}$). Only the linear polarization accommodates singularities (instantaneous zero field vectors; blue dot) at the instant $\mathrm{t=T_{2}}$.} \label{fig1}
\end{figure} 

For an optical scatterer, its optical chirality can be characterized by a generalized circular dichroism parameter $\mathbb{CD}$ defined as:

\begin{equation}
\label{CD-original}
\mathbb{CD}(\mathbf{k}_{i})=(\rm {C}^{\circlearrowright}-\rm {C}^{\circlearrowleft})/(\rm {C}^{\circlearrowright}+\rm {C}^{\circlearrowleft}),
\end{equation}
where $\mathbf{k}_{i}$ denotes the incident direction vector; $\mathrm{C}$ is a cross section (of either extinction, scattering or absorption~\cite{Bohren1983_book}): its superscript denotes the handedness of incident wave; its dependence  on $\mathbf{k}_i$ is suppressed.  Generally the value of $\mathbb{CD}$ would be dependent on the specific type of cross section chosen, \textit{e.g.} $\mathbb{CD}$ calculated through scattering cross section would be distinct from that through absorption cross section~\cite{CHEN_2020_Phys.Rev.Research_Scatteringa}. While in the single-mode regime, such dependence would be eliminated and $\mathbb{CD}$ would be invariant for all three types of cross sections~\cite{CHEN_Phys.Rev.Lett._Extremize}. So our central question can be reformulated as: is it possible that $\mathbb{CD}(\mathbf{k}_{i})\neq0$ for all $\mathbf{k}_{i}$ throughout the momentum sphere?

The single mode dominantly supported would radiate to all directions (the radiation direction is denoted by $\mathbf{k}_{r}$) and along each direction the radiation polarization (polarization for the radiated transverse far field) is characterized partially by the third Stokes parameter $\mathbb{S}_3$ (signed ellipticity)~\cite{Bohren1983_book}: $\mathbb{S}_3(\mathbf{k}_{r})=\pm1$ corresponds respectively to RCP and LCP radiations along $\mathbf{k}_{r}$; $\mathbb{S}_3(\mathbf{k}_{r})=0$ corresponds to linearly-polarized radiations. If the scatterer is reciprocal, it has been proved that~\cite{CHEN_Phys.Rev.Lett._Extremize}:
\begin{equation}
\label{CD}
\mathbb{CD}(\mathbf{k}_{i})=\mathbb{S}_3(\mathbf{k}_{r}=-\mathbf{k}_{i}),
\end{equation}
which means that $\mathbb{CD}$ is solely decided by the mode radiation polarization opposite to the incident direction. Here a special scenario merits special attention: there might exist a  direction ($\mathbf{k}_{0}$) along which the radiation is zero and thus the polarization is ill-defined. For waves incident opposite to this direction ($\mathbf{k}_{i}=-\mathbf{k}_{0}$),  reciprocity secures that, irrespective of the incident polarization, the mode would not be excited. More specifically, $\rm {C}^{\circlearrowright}=\rm {C}^{\circlearrowleft}=0$ in Eq.~(\ref{CD-original}) and thus $\mathbb{CD}(\mathbf{k}_{i}=-\mathbf{k}_{0})$ is also ill defined. Nevertheless, with $\mathbf{k}_{i}=-\mathbf{k}_{0}$, since the scatter would be excited by neither RCP nor LCP waves, the optical response is identical and thus reasonable to set $\mathbb{CD}(\mathbf{k}_{i}=-\mathbf{k}_{0})=0$. To ensure the validity of Eq.~(\ref{CD}) for the special zero-radiation direction, we accordingly also set $\mathbb{S}_3(\mathbf{k}_{0})=0$.

According to  Eq.~(\ref{CD}), the possibility of  $\mathbb{CD}(\mathbf{k}_{i})\neq0$ for all $\mathbf{k}_{i}$ is equivalent to another possibility of  $\mathbb{S}_3(\mathbf{k}_{r})\neq0$ for  all $\mathbf{k}_{r}$. As has been argued,  $\mathbb{S}_3(\mathbf{k}_{r})=0$ corresponds to zero or linearly-polarized radiations, and thus our central question is essentially mapped into the following: for the single mode supported, is it possible that along all directions its radiations are neither zero nor linearly polarized?

A simple topological argument can easily rule out the aforementioned possibility~\cite{WEN_2025__Polarizations}, as illustrated in Fig.~\ref{fig1}. The radiations of the mode are described by not only polarization ellipses [dashed lines in Figs.~\ref{fig1}(b) and \ref{fig1}(c)] but also cyclically rotating (electric or magnetic) field vectors tracing out them [arrows in Figs.~\ref{fig1}(b) and \ref{fig1}(c)].  Those vectors correspond to continuous tangent vector fields on the momentum sphere [Fig.~\ref{fig1}(a)]. Though at different instants the vector field distributions are distinct (the field vectors are cyclically rotating), at any instant  the Poincaré-Hopf theorem (of which a reduced scenario is the hairy ball theorem)~\cite{NEEDHAM__Visuala} requires that there must exist isolated directions where the instantaneous vectors are zero (singular) [Figs.~\ref{fig1}(a) and \ref{fig1}(b)]. For elliptical and circular polarizations, the field vector is never zero at any instant [Fig.~\ref{fig1}(c)], and consequently the vectorial singularities shown in Fig.~\ref{fig1}(a) can only be found at directions of zero radiation ($\mathbf{k}_{0}$) or linearly-polarized radiations ($\mathbf{k}_{\mathrm{L}}$).  That is, the Poincaré-Hopf theorem prohibits $\mathbb{S}_3(\mathbf{k}_{r})\neq0$ for  all $\mathbf{k}_{r}$, and thus also prohibits $\mathbb{CD}(\mathbf{k}_{i})\neq0$ for all $\mathbf{k}_{i}$ according to Eq.~(\ref{CD}).  This brings us to our central conclusion: for a reciprocal single-mode scatter, there must exist some incident directions of optical achirality. 

\begin{figure}[tp]
	\centerline{\includegraphics[width=0.4\textwidth]{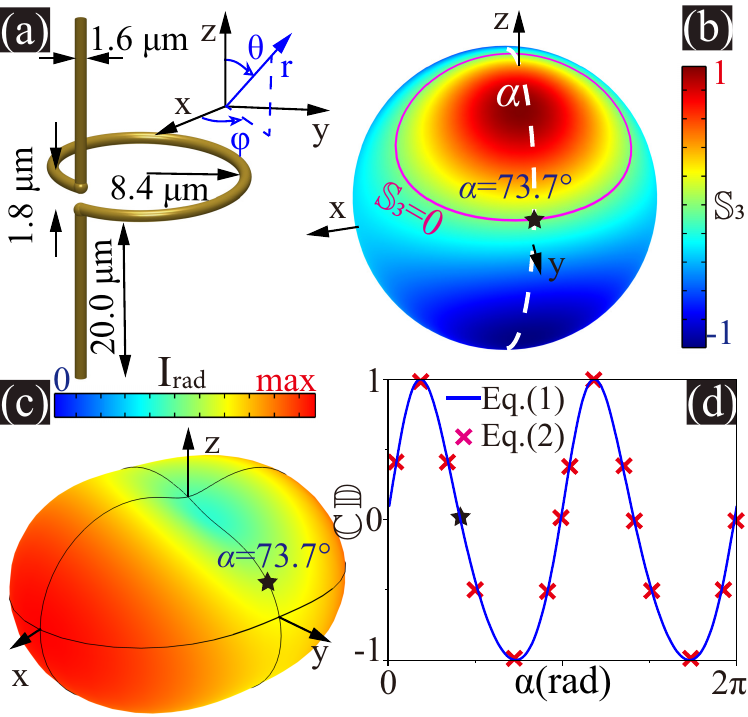}} \caption{\small (a) A reciprocal chiral gold particle with a spherical coordinate system ($r,~\theta,~\varphi$) also included.  (b) $\mathbb{S}_3$ distributions for the far-field QNM radiations. Directions of linearly-polarized radiations ($\mathbf{k}_{\mathrm{L}}$) are marked by an enclosed circuit (magenta) and another such circuit of $\mathbf{k}_{\mathrm{L}}$ is out of view (on the other side of the momentum sphere shown).  A great circle on the plane of $\varphi=90^\circ$ and $270^\circ$ (marked by dashed lines; parametrized by $0 \leq \alpha \leq 2\pi$) is also shown, on which one of four directions of linear polarizations [the great circle intersects with the two circuits of linear polarizations at four points, where $\mathbb{CD}=0$ in (d)]  is indicated (black $\star$; $\alpha=73.7^\circ$). (c) Angular distributions of radiation intensity for the QNM. (d) Angular $\mathbb{CD}$ spectra for waves incident antiparallel to directions on the great circle of directions marked in (b).}
\label{fig2}
\end{figure} 

Now we proceed to specific structures to exemplify our discovery. We begin with a representative chiral particle [two identical circular cylinders (radius $0.8~\mu$m and length $20~\mu$m) connected by a circular helix segment (radius $8.4~\mu$m) with one pitch ($1.8~\mu$m)] shown in Fig.~\ref{fig2}(a), with all geometric parameters specified . The particle is reciprocal and made of gold with experimental relative permittivity~\cite{Johnson1972_PRB} fitted by a Drude model: $\epsilon_r(\omega)=1-\omega_{p}^{2}/(\omega^2+i\Gamma\omega)$ with plasma frequency $\omega_p=1.37\times10^{16}~\mathrm{rad/s}$ and collision frequency $\Gamma=8.17\times10^{13}~\mathrm{rad/s}$. Throughout this paper, all numerical results are obtained through the commercial software COMSOL Multiphysics.  For such a chiral particle, a spectrally isolated QNM is identified with complex eigenfrequency $\widetilde{\omega}_{\rm{A}}=(8.18\times 10^{12}+1.62\times10^{11}\textit{i})~\mathrm{rad/s}$  (its real part corresponds to the vacuum wavelength ${\lambda}_{\rm{A}}=230.2~\mu$m). The $\mathbb{S}_3$ distributions for the far-field QNM radiations are shown in Fig.~\ref{fig2}(b), where directions of linear polarizations ($\mathbf{k}_{\mathrm{L}}$ along which $\mathbb{S}_3=0$) are marked by closed solid lines (codimension analysis reveals that on the momentum sphere the linear polarizations form closed lines~\cite{BERRY_2004_J.Opt.PureAppl.Opt._Index}; another linear polarization circuit is out of view). One direction of linear polarization is marked (black $\star$; $\theta=73.7^\circ$ and $\phi=90^\circ$), and a great circle passing this direction on the momentum sphere (parameterized by $0\leq\alpha\leq2\pi$; $\alpha=0$ and $\pi$ correspond respectively to +\textbf{z} and -\textbf{z} directions) is also indicated (dashed lines) in Fig.~\ref{fig2}(b). Radiation intensity ($\mathbf{I}_{\mathrm{rad}}$ as the magnitude of Poynting vector or far field intensity) distributions of the QNM are shown in Fig.~\ref{fig2}(c),  for which there are no directions of zero radiations ($\mathbf{I}_{\mathrm{rad}}\neq0$ for all $\mathbf{k}_{r}$). The angular $\mathbb{CD}$ spectra for plane waves incident antiparallel to the directions on the great circle are shown in Fig.~\ref{fig2}(d), with the incident wavelength fixed at ${\lambda}_{\rm{A}}$. For the  $\mathbb{CD}$ spectra, we show two sets of results [obtained according to Eqs.~(\ref{CD-original}) and (\ref{CD})], which agree perfectly with each other. As is clearly shown, $\mathbb{CD}=0$ at the four points of linear polarizations (where the great circle interacts with the two circuits of linear polarizations; only two of them are in view and one is marked by $\star$).

\begin{figure}[tp]
	\centerline{\includegraphics[width=0.4\textwidth]{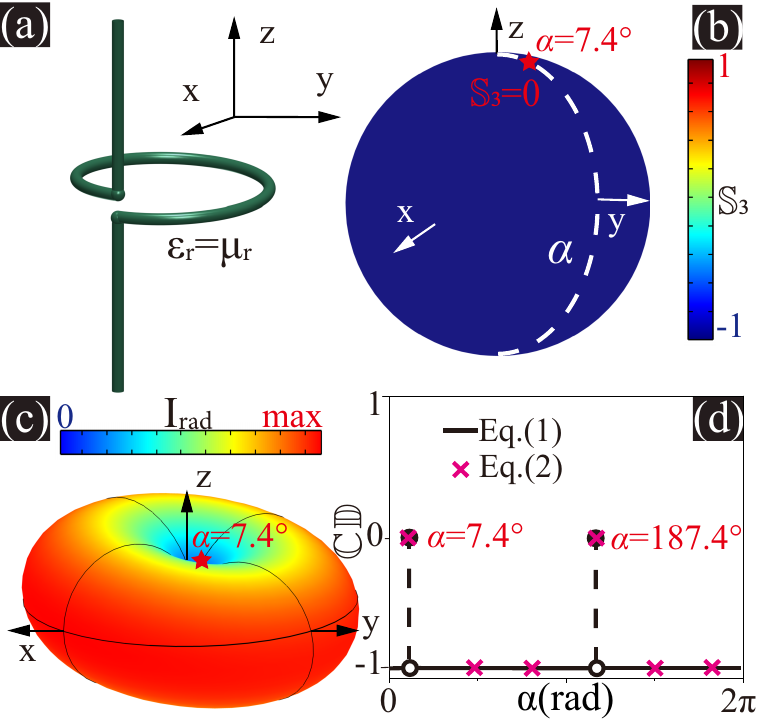}} \caption {\small (a) A self-dual reciprocal chiral particle with geometric parameters being identical to those specified  in Fig.~\ref{fig2}(a). (b) $\mathbb{S}_3$ distribution for the QNM radiations: $\mathbb{S}_3=-1$ for all directions, excpet for two antiparrael directions (one of them is in view and marked) where the radiations are zero ($\mathbb{S}_3=0$). (c) Radiation intensity distributions of  the QNM, and ther radiation is zero along $\alpha=7.4^\circ$ and its antiparallel direction $\alpha=187.4^\circ$. (d) Angular $\mathbb{CD}$
spectra for waves incident antiparallel to directions on a great circle of directions marked in (b), where $\mathbb{CD}=0$ for the two directions of zero radiations.}
\label{fig3}
\end{figure} 

The optical achirality ($\mathbb{CD}=0$) revealed above is induced by directions of linear polarizations. When such directions are absent, as has been already argued, there then must be directions of zero radiations ($\mathbf{k}_{0}$ where $\mathbf{I}_{\mathrm{rad}}=0$) that can also secure optical achirality. It is worth mentioning that directions of zero radiations are non-generic and structurally unstable~\cite{BERRY_2004_J.Opt.PureAppl.Opt._Index}: tiny perturbations would break such points into pairs of  directions of circular polarizations with opposite handedness, which are separated by lines of linear polarizations. As a result, lines of linear polarizations and points of circular polarizations are generic and structurally stable upon perturbations, as is observed in Fig.~\ref{fig2}(b) [circular polarizations locate at the positions where $\mathbb{CD}$ reach its extreme values of $\mathbb{CD}=\pm1$ in Fig.~\ref{fig2}(d)]. To exclude the presence of linear polarizations and thus ensure the presence of zero radiations, extra special symmetries are required, of which an outstanding example is the electromagnetic duality symmetry~\cite{WEN_2025__Polarizations,FERNANDEZ-CORBATON_2013_Phys.Rev.Lett._Electromagnetica}. 

As a next step, we turn to an ideal self-dual particle ($\epsilon_r=\mu_r=36$, where $\mu_r$ is the relative permeability) schematically shown in Fig.~\ref{fig3}(a), for which the geometric parameters are the same as those specified in Fig.~\ref{fig2}(a), and an isotropic $\mu_r$ makes sure that the structure is still reciprocal~\cite{CALOZ_2018_Phys.Rev.Applied_Electromagnetic}. Similarly, one individual QNM supported by the particle is identified with complex eigenfrequency $\widetilde{\omega}_{\rm{B}}=(2.35\times 10^{13}+2.97\times10^{11}\textit{i})~\mathrm{rad/s}$ 
 (${\lambda}_{\rm{B}}=80.18~\mu$m). Its corresponding distributions of $\mathbb{S}_3$ and  $\mathbf{I}_{\mathrm{rad}}$ are shown respectively in Figs.~\ref{fig3}(b) and \ref{fig3}(c). It is clear from Fig.~\ref{fig3}(b) that $\mathbb{S}_3(\mathbf{k}_{r})=-1$ for all $\mathbf{k}_{r}$ (as required by the duality symmetry), except for two directions where $\mathbf{I}_{\mathrm{rad}}=0$ (as required by the global topology). In other words, there are no directions of linear polarizations.  One direction of zero radiation is marked ($\theta=7.4^\circ$ and $\phi=90^\circ$), and a great circle passing this point on the momentum sphere is also indicated  in Fig.~\ref{fig2}(b). The angular $\mathbb{CD}$ spectra for plane waves incident antiparallel to the directions on the great circle are shown in Fig.~\ref{fig2}(d), with the incident wavelength fixed at ${\lambda}_{\rm{B}}$. As is clearly shown, $\mathbb{CD}=0$ at the two points of zero radiations, and at all other points $\mathbb{CD}$ is extremized to be $\mathbb{CD}=-1$.

To conclude,  we have merged the mathematical theorem of Poincaré-Hopf, the physical principle of reciprocity, and the concept of instantaneous vector field singularity, to provide a partial answer to our central question: \textit{Is it possible for a structure to be optically chiral for all incident directions?} We prove that this is not possible for a single-mode reciprocal structure, for which there must exist directions of topology protected optical achirality. The conclusion we draw is widely applicable, irrespective of not only the structural geometric parameters and material optical parameters, but also the wavelength of incident waves.  For general nonreciprocal and/or multi-mode structures, we have not managed to get a comprehensive answer and further explorations would benefit not only the obviously relevant fields of chiral optics and topological photonics, but also other branches of electromagnetic waves and general physical waves, where singularity and topology are prevalent.  \\

\section*{acknowledgement}
This research was funded by the National Natural Science Foundation of
China (12274462, 11674396, and 11874426) and several other projects of Hunan Province (2024JJ2056, 2023JJ10051), 2018JJ1033 and 2017RS3039). 




\begin{thebibliography}{10}
\newcommand{\enquote}[1]{``#1''}

\bibitem{BARRON_2009__Molecular}
L.~D. Barron, \emph{Molecular {{Light Scattering}} and {{Optical Activity}}}
  ({Cambridge University Press}, 2009).

\bibitem{MUN_LightSci.Appl._Electromagnetic}
J.~Mun, M.~Kim, Y.~Yang, T.~Badloe, J.~Ni, Y.~Chen, C.-W. Qiu, and J.~Rho,
  \enquote{Electromagnetic chirality: From fundamentals to nontraditional
  chiroptical phenomena,} Light Sci. Appl. \textbf{9}, 139 (2020).

\bibitem{CHEN_Phys.Rev.Lett._Extremize}
W.~Chen, Q.~Yang, Y.~Chen, and W.~Liu, \enquote{Extremize {{Optical
  Chiralities}} through {{Polarization Singularities}},} Phys. Rev. Lett.
  \textbf{126}, 253901 (2021).

\bibitem{LALANNE__LaserPhotonicsRev._Light}
P.~Lalanne, W.~Yan, K.~Vynck, C.~Sauvan, and J.-P. Hugonin, \enquote{Light
  {{Interaction}} with {{Photonic}} and {{Plasmonic Resonances}},} Laser
  Photonics Rev. \textbf{12}, 1700113 (2018).

\bibitem{Bohren1983_book}
C.~F. Bohren and D.~R. Huffman, \emph{Absorption and Scattering of Light by
  Small Particles} (Wiley, 1983).

\bibitem{CHEN_2020_Phys.Rev.Research_Scatteringa}
W.~Chen, Q.~Yang, Y.~Chen, and W.~Liu, \enquote{Scattering activities bounded
  by reciprocity and parity conservation,} Phys. Rev. Research \textbf{2},
  013277 (2020).

\bibitem{WEN_2025__Polarizations}
C.~Wen, J.~Zhang, C.~Zhang, S.~Qin, Z.~Zhu, and W.~Liu, \enquote{Polarizations
  {{Underdescribe Vectorial Electromagnetic Waves}},} arXiv:2406.06132  (2025).

\bibitem{NEEDHAM__Visuala}
T.~Needham, \emph{Visual {{Differential Geometry}} and {{Forms}}: {{A
  Mathematical Drama}} in {{Five Acts}}} ({Princeton University Press},
  {Princeton}, 2021).

\bibitem{Johnson1972_PRB}
P.~B. Johnson and R.~W. Christy, \enquote{Optical constants of the noble
  metals,} Phys. Rev. B \textbf{6}, 4370 (1972).

\bibitem{BERRY_2004_J.Opt.PureAppl.Opt._Index}
M.~V. Berry, \enquote{Index formulae for singular lines of polarization,} J.
  Opt. A: Pure Appl. Opt. \textbf{6}, 675 (2004).

\bibitem{FERNANDEZ-CORBATON_2013_Phys.Rev.Lett._Electromagnetica}
I.~{Fernandez-Corbaton}, X.~{Zambrana-Puyalto}, N.~Tischler, X.~Vidal, M.~L.
  Juan, and G.~{Molina-Terriza}, \enquote{Electromagnetic {{Duality Symmetry}}
  and {{Helicity Conservation}} for the {{Macroscopic Maxwell}}'s
  {{Equations}},} Phys. Rev. Lett. \textbf{111}, 060401 (2013).

\bibitem{CALOZ_2018_Phys.Rev.Applied_Electromagnetic}
C.~Caloz, A.~Al{\`u}, S.~Tretyakov, D.~Sounas, K.~Achouri, and Z.-L.
  {Deck-L{\'e}ger}, \enquote{Electromagnetic {{Nonreciprocity}},} Phys. Rev.
  Applied \textbf{10}, 047001 (2018).

\end{thebibliography}

\end{document}